\newif\ifonecol 
\onecolfalse 

\ifonecol
\documentclass[journal,comsoc,12pt,onecolumn]{IEEEtran}
\else
\documentclass[journal,comsoc]{IEEEtran}
\fi

\newlength{\figurewidth}
\ifonecol
\else
\fi

\ifonecol
\usepackage{setspace}
\fi

\usepackage{ifpdf}
\ifpdf
\fi

\usepackage{graphicx}
\usepackage{afterpage}
\usepackage{empheq}
\usepackage{amsmath}
\usepackage{amsfonts}
\usepackage{amsbsy}
\usepackage{amssymb}
\usepackage{cite}
\usepackage{bm}
\ifonecol
\usepackage[varg]{txfonts}
\let\mathbb=\varmathbb
\DeclareSymbolFont{letters}{OML}{ztmcm}{m}{it}
\fi
\usepackage[all]{xy}
\usepackage{subfigure}
\usepackage{floatrow}
\usepackage{color}
\usepackage{xcolor}
\usepackage{enumerate}
\newcommand{\MYfooter}{\smash{\scriptsize
\hfil\parbox[t][\height][t]{\textwidth}{\centering
\copyright 2019 IEEE. Personal use of this material is permitted. Permission from IEEE must be obtained for all other uses, including reprinting/republishing this material for advertising or promotional purposes, collecting new collected works for resale or redistribution to servers or lists, or reuse of any copyrighted component of this work in other works. DOI: 10.1109/TCOMM.2018.2870874.}\hfil\hbox{}}}

\makeatletter

\def\ps@IEEEtitlepagestyle{%
\def\@oddfoot{\MYfooter}%
\def\@evenfoot{\MYfooter}}

\makeatother

\begin{document}
%
\title{Chaos-Based Multicarrier VLC Modulator with Compensation of LED Nonlinearity}
%
%
%

\author{Francisco J. Escribano,~\IEEEmembership{Senior Member,~IEEE,} and Jos\'e S\'aez-Landete and Alexandre Wagemakers
\thanks{Francisco J. Escribano and Jos\'e S\'aez-Landete are with the Nonlinear Communications and Signal Processing Research Group,
Universidad de Alcal\'a, 28805 Alcal\'a de Henares, Spain (email: francisco.escribano@ieee.org, jose.saez@uah.es).}
\thanks{Alexandre Wagemakers is with the Nonlinear Dynamics and Chaos Theory Group, Universidad Rey Juan Carlos, 28933 M\'ostoles, Spain (email: alexandre.wagemakers@urjc.es).}}

%
%

\markboth{Journal of \LaTeX\ Class Files,~Vol.~XX, No.~Y, February~2018}{Shell \MakeLowercase{\textit{et al.}}: Bare Demo of IEEEtran.cls for IEEE Communications Society Journals}
%



\maketitle

\begin{abstract}
The massive deployment of LED lightning infrastructure has opened the opportunity to reuse it as Visible Light Communication (VLC) to leverage the current RF spectrum crisis in indoor scenarios. One of the main problems in VLC is the limited dynamic range of LEDs and their nonlinear response, which may lead to a severe degradation in the communication, and more specifically in bit error rate (BER). This is aggravated by the extensive usage of multicarrier multiplexing, based on optical Orthogonal Frequency Disivion Multiplexing (O-OFDM), characterized by a high Peak-to-Average Power Ratio (PAPR). Here we present a chaos-based coded modulation (CCM) setup specifically adapted to the LED nonlinearities. It replaces the usual modulation, while keeping the multicarrier O-OFDM structure unchanged. First, we obtain a semi-analytical bound for the bit error probability, taking into account the LED nonlinear response. The bound results particularly tight for the range of signal-to-noise ratio of interest. Then, we propose a method to design the modulator based on optimization techniques. The objective function is the semi-analytical bound, and the optimization is applied to a parametrization of the CCM conjugation function. This appropriately shapes the chaotic waveform, and leads to BER improvements that outperform classical counterparts under ideal predistortion.
\end{abstract}


%
\IEEEpeerreviewmaketitle

\section{Introduction}
%
%
%
%

\IEEEPARstart{R}{ecently}, new communication systems outside the Radio Frequency (RF) domain are being proposed to alleviate the growing demand for bandwidth in the current context of wireless spectrum shortage. Some new proposals are based on exploiting the optical domain, and are labeled under the terms Optical Wireless Communications (OWC), in general, or Visible Light Communications (VLC), in particular, when dealing exclusively with the visible range. This latter alternative is attractive because of the potential reuse factor of existing light-emitting diode (LED) lighting infrastructure. Moreover, optical-based communication technologies have other advantages such as low cost, energy efficiency, low electromagnetic interference (EMI) and security in the communications \cite{Dimitrov15}.

In almost every communication system, the nonlinear behavior and the constraints in the dynamic range of some electrical components are two of the main causes of the degradation of its theoretical performance. In RF communications, the power amplifier, digital-to-analog and analog-to-digital converters are examples of these components. In OWC and, most recently, in VLC, the LED working in the transmitter has proven to be the most determinant nonlinear component \cite{elgala2009non}. On the one hand, the LED has a relatively reduced dynamic range, and this generates a potentially destructive double-sided clipping in the transmitted signal. On the other hand, the LED exhibits nonlinear behavior both in the voltage-current and in the current-light intensity relationship, and this determines a nonlinear response within its dynamic range \cite{ying2015nonlinear}.

In modern communication systems, be them RF-based or optical-based, Orthogonal Frequency Division Multiplexing (OFDM) has been imposed as the preferred medium access and signal generation technique due to its ability to handle high data rates, its superior spectral efficiency, and the simplicity in implementing channel equalization. One of the costs for all these advantages is the resulting large dynamic range of the transmited signal. Therefore, the signal has a high Peak-to-Average Power Ratio (PAPR), which may severely affect the overall performance of the system. Given the said advantages, OFDM has also become popular in OWC and VLC. The problem is that here the limitations in linearity and dynamic range are more demanding than in RF \cite{Dimitrov15}. This situation is more critical in VLC, where the LED has lighting as its primary function. Different strategies to mitigate the LED nonlinear distortion and signal clipping have been proposed so far, and a general description of them can be seen in \cite{ying2015nonlinear}. In general, the strategies can be classified into two large groups. The first group is based on waveform shaping to reduce the signal PAPR, which indirectly reduces the effect of the limited dynamic range and nonlinear behavior of the LED response. One example it can be seen in \cite{doblado2015cubic}.

The second group is based on nonlinearity correction. If it takes place at the transmitter is named predistortion \cite{dimitrov2013information}, whereas if it takes place at the receiver is named postdistortion \cite{qian2014adaptive}. Some strategies require a reliable LED nonlinear model, for which, in turn, there are two possible approximations. The first and simplest is a memoryless model, consisting in a transfer function modeled by a polynomial \cite{dimitrov2013information}. If the frequency-dependent response of the LED is considered, the complexity increases and a model with memory is required. In this case, the Volterra model is the most popular \cite{kamalakis2011empirical}. Other strategies do not require an explicit LED model, like the so-called reproducing kernel Hilbert space (RKSH) \cite{Mitra17}, or the Wiener (a subset of Volterra) or Hammerstein models \cite{qian2014adaptive,ying2015nonlinear}, which in many cases are applied to blind postdistortion. The predistortion approximation is usually more interesting, because trough scaling and biasing, the signal can be optimally conditioned, taking into account the dynamic range constraints and nonlinear response of the LED, as shown in \cite{dimitrov2013information} and \cite{ying2015optimization}.

In this paper, a different approximation is developed. Based on the memoryless nonlinear LED model, we propose to optimize the codification of the information through a chaos-based coded modulator (CCM) \cite{Escribano09}. Other coded modulation schemes, like the trellis coded modulation (TCM) based ones, have been considered in nonlinear transmission schemes \cite{Montezuma10,Alreesh16}, or even for OWC \cite{Fath12,Wang13}. In general, chaos-based communication systems, acting at the waveform or the coding level (or both), are already part of the state-of-the-art in communications \cite{Kaddoum16}. In the specific case of CCM systems, their inherent nonlinear nature and their coded structure have made them well fitted to the nonlinear response of the power amplifiers in RF \cite{Escribano16b}, and to other kind of nonlinear channels \cite{Escribano08a}. Moreover, their TCM-like background has made them very easy to implement in competitive conditions \cite{Wagemakers12, Wagemakers17}, unlike other chaotic inspired systems. We show there how the CCM inherent ability to adapt its coded modulated nature to the nonlinear response of the LED leads to a noticeable improvement in the error rate of the system under nonlinear adverse conditions. This improvement relies on the fact that chaotic trajectories and the specific dynamics of the chaos-based communication system can be conveniently tuned by means of an appropriate conjugation function \cite{Baranovsky95,Escribano06a}. The CCM overall response can be fully designed by means of an optimization procedure, which allows to obtain the conjugation function which minimizes a semi-analytical bound for the bit error probability, based on the pairwise error probability (PEP) \cite{Escribano16b}.

For these purposes, the rest of the paper is organized as follows. Section \ref{setup} will be devoted to the description of the OWC-CCM system, and its setup from the point of view of the required analysis previous to the calculation of the optimization cost function. Section \ref{analysis} will be devoted to the derivation of the bit error probability function to be minimized. Section \ref{optim} gives the details about the optimization procedure and its feasibility. Section \ref{results} presents the results and their corresponding discussion. Section \ref{implementation} clarifies some aspects about the possible implementation of the system. Finally, Section \ref{conclusions} is devoted to the conclusions and the description of the future work.

\section{System setup}
\label{setup}

\ifonecol
\begin{figure}
\centering
\includegraphics[width=\textwidth, keepaspectratio]{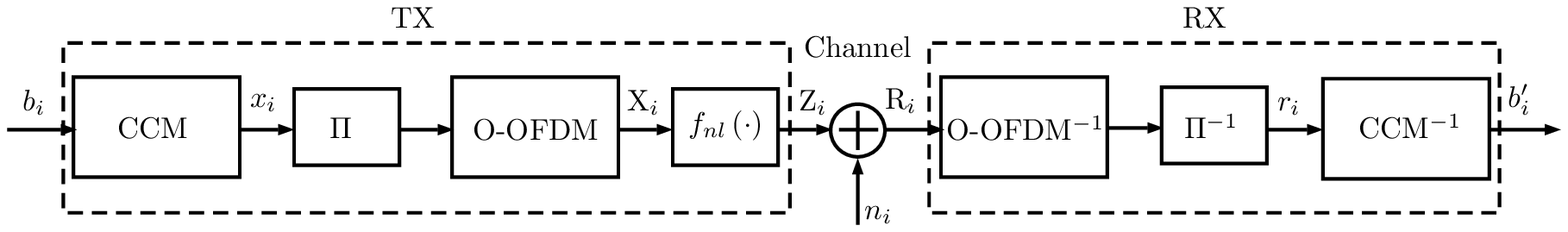}
\caption{\label{fig:BlockDiagram} Block diagram of the model for the communication system.}
\end{figure}
\else
\begin{figure*}
\centering
\includegraphics[width=\textwidth, keepaspectratio]{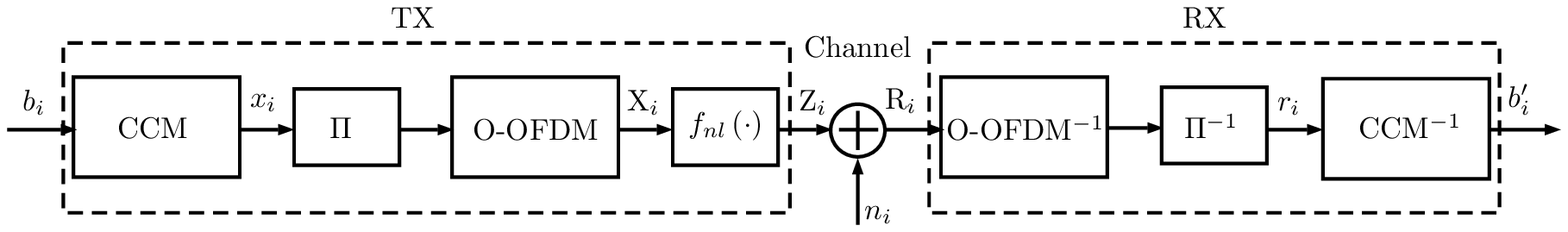}
\caption{\label{fig:BlockDiagram} Block diagram of the model for the communication system.}
\end{figure*}
\fi
In Figure \ref{fig:BlockDiagram} we represent the scheme of the model of the communication system considered for the VLC setup of this article, including the transmitter, the channel and the receiver. We follow the approach of \cite{Dimitrov13} in order to ease analysis and simulation, as will be made evident in the sequel. In this first approach, we consider the context of a point-to-point transmission between a static VLC transmitter and a static VLC receiver. The transmitter is fed with an independent and identically distributed (iid) binary sequence $b_i$, in consecutive blocks of $M$ bits. Each block is then processed to produce a chaos-based sequence $x_i$ of length $M$.
The chaos-based coded modulators (CCMs) considered here are of the kind described in \cite{Escribano09} and \cite{Escribano14}, and we review their principles here for clarification. This type of chaotic encoders are in fact chaotic maps controlled by small perturbations as proposed in \cite{Kozic06a}. They are described by a recursion in the form
\begin{equation}
\label{eqn1}
z_i=f\left( z_{i-1},b_i \right)+b_i \cdot  2^{-Q},
\end{equation}
where chaotic applications $f\left( \cdot, 0 \right)$ and $f\left( \cdot, 1 \right)$ leave the interval $\left[0,1\right]$ invariant. In addition, they are piecewise linear with slope $\pm 2$ wherever it is defined. The natural number $Q$ is the number of bits used to represent $z_i$. The recursion (\ref{eqn1}) leaves a finite set invariant, and therefore we can restrict it to $z_i \in S_Q=\{ m \cdot 2^{-Q} | m=0,1, \cdots, 2^{Q}-1 \}$. When $Q \rightarrow \infty$, expression (\ref{eqn1}) becomes simply the recursion by the chaotic maps, depending on the value of $b_i$. Note that these chaotic encoders are intrinsically non-systematic.

We may consider different pairs of applications $f\left( \cdot, 0 \right)$ and $f\left( \cdot, 1 \right)$, as seen in\cite{Escribano09}, but we restrict here to the so-called multi-tent map (mTM):
\begin{eqnarray}
\label{eqn5}
f(z,0) & = & 1-|2z-1|, \nonumber \\
f(z,1) & = & 3/2-|2z-1| \mod 1.
\end{eqnarray}
The systems described by (\ref{eqn1}) can be represented by means of a recursive convolutional encoder and a mapping to the signal constellation given by $S_Q$. Thus, the system is in fact equivalent to a trellis coded modulation (TCM) \cite{Schlegel04}, with $2^Q$ states and requiring $Q$ binary memory positions. The equivalent finite-state recursive convolutional encoder is given by
\begin{eqnarray}
\label{eqn6}
v^i_Q & = & u_1 v^{i-1}_{Q} \oplus u_2 v^{i-1}_{Q-1} \oplus u_3 b_i, \nonumber \\
v^i_j & = & u_4 v^{j-1}_{i-1} \oplus u_5 v^{i-1}_{Q}, \, \, \,\, j=Q-1, \cdots, 2, \nonumber \\
v^i_1 & = & u_6 b_1.
\end{eqnarray}
where $v^i_j$ are contents of the $j-$th memory position at time $i$, and $\mathbf{u} = (u_1,u_2, \cdots, u_6)$ is a binary vector. The mapping to the signal constellation is given by the expression $z_i = \sum_{i=1}^{Q} 2^{-(Q+1-i)} v_i$. The vector $u=\left[1,1,1,1,1,1\right]$ corresponds to the recursion of the mTM. Other vectors $\mathbf{u}$ define other kinds of CCM. Memory positions are initialized to state $\mathbf{0}$ for convenience.

The $z_i$ samples produced by this process are subject to a non-decreasing conjugation function
\begin{equation}\label{eq:fconj}
s_i=g\left(z_i\right)
\end{equation}
mapping the interval $\left[0,1\right]$ into itself \cite{Escribano06a}. The sequence $s_i$ is then mapped to a phase constellation as in \cite{Escribano14}, following
\begin{equation}
x_i=\mathrm{e}^{2\pi j s_i}.
\end{equation}
Given the symmetry present in the CCM infrastructure and the conjugation function, $x_i$ constitutes a zero-mean sequence, with average power $\sigma_x^2=1$. The conjugation function $g\left(\cdot\right)$ will be parametrized by means of a set of independent variables subject to certain constraints, and subsequently optimized in order to compensate the effects on the BER of the LED nonlinearity, as in \cite{Escribano16b}, where the same concepts were applied in the context of RF high power amplifier (HPA) nonlinearity.

At this point, the block of $M$ output samples is interleaved, and the result feeds an optical OFDM modulator (O-OFDM). The interleaver is a pseudorandom one, acting on successive blocks of length $M$. Unlike a standard OFDM modulator, the optical OFDM modulator must generate a signal that has to be real and positive, in order to drive the LED. There are different approaches to achieve this with a multicarrier signal, and one of the most popular consists in forcing Hermitian symmetry in the symbols in order to obtain a real signal. After this, an appropriate bias is added in order to limit the lower clipping. This method is named DC Biased Optical-Orthogonal Frequency Division Modulation (DCO-OFDM) \cite{Gonzalez06}. We use this approach to create the multicarrier output sequence $\mathrm{X}_i$, with a particularity: $\mathrm{X}_i$ is just the result of applying to $x_i$ the Hermitian extension and then the IFFT. For reasons that will be made evident in the analysis, the bias factor is incorporated in the overall nonlinear transfer function $f_{nl}\left( \cdot \right)$ as in \cite{Dimitrov13}. This fact is shown in the accompanying Figure \ref{fig:OFDMDiagram}.

If the order of the IFFT applied at the O-OFDM block is $N$, and taking into account the Hermitian extension performed, there will be $M/(N/2-1)$ output OFDM symbols. For convenience and without loss of generality, $M$ is chosen to fit an integer number of OFDM symbols of order $N$. We assume that the length of the cyclic prefix added is enough to compensate for the delay spread of the channel, and subsequently we do not include explicitly said prefix in the system model, and we do not take into account any intersymbol interference (ISI) distortion. Symbols $\mathrm{X}_i$ go through the nonlinear function $f_{nl}\left( \cdot \right)$, as depicted in Figure \ref{fig:BlockDiagram}, which includes the nonlinear transfer function of the LED together with the biasing and de-biasing operations \cite{Dimitrov13}, producing
\begin{equation}
\label{eq0}
\mathrm{Z}_i=f_{nl}\left( \mathrm{X}_i \right)=F_{nl} \left( \rho \cdot \mathrm{X}_i + \beta_{DC} \right) - F_{nl} \left( \beta_{DC} \right),
\end{equation}
where $\beta_{DC}$ is the input bias to limit lower clipping in the signal, $F_{nl} \left( \cdot \right)$ is the real-valued LED nonlinear transfer function, $F_{nl} \left( \beta_{DC} \right)$ represents the output de-biasing, and $\rho$ is a back-off factor to control the severity of the clipping and nonlinear effects. Real de-biasing would happen while demultiplexing at the receiver, but, for convenience of the analysis and without loss of generality, we consider it at this stage. In fact, since we do not take into account the DC value in the computation of the signal-to-noise ratio, this does not affect the results (as will be made evident in the analytic part). Equality throughout all the comparisons is guaranteed since we do the same with all the alternatives. The input back-off factor IBO is defined as
\begin{equation}
\mathrm{IBO}\left( \mathrm{dB} \right)=-10 \cdot \log_{10} \left( \rho^2 \cdot \mathrm{E}\left[ \mathrm{X}_i^2 \right] \right).
\end{equation}
The model for $F_{nl}\left(\cdot\right)$ is taken normally as a symmetric function with upper and lower clipping levels, and a polynomial evolution between both. The details of this model and how it relates to the physical one can be seen in Figure \ref{fig:OFDMDiagram}.
\ifonecol
\begin{figure}
\centering
\includegraphics[width=0.7\textwidth, keepaspectratio]{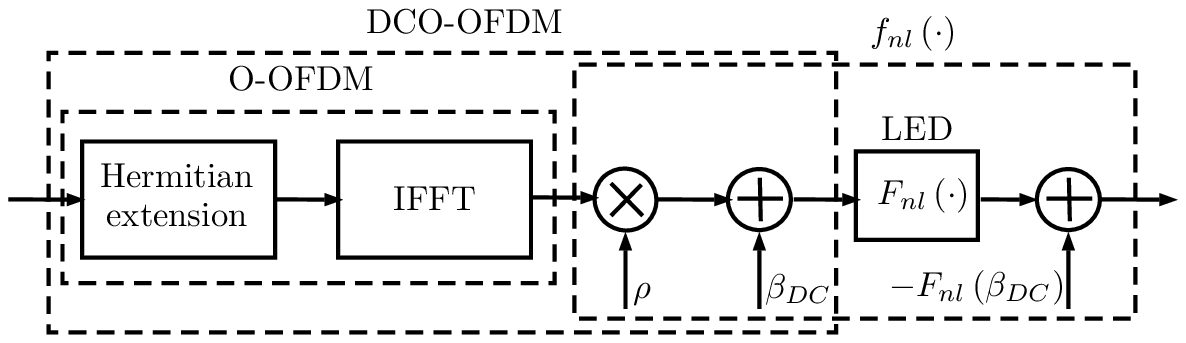}
\caption{\label{fig:OFDMDiagram} Block diagram showing how the physical model is handled for analysis.}
\end{figure}
\else
\begin{figure}
\centering
\includegraphics[width=\figurewidth, keepaspectratio]{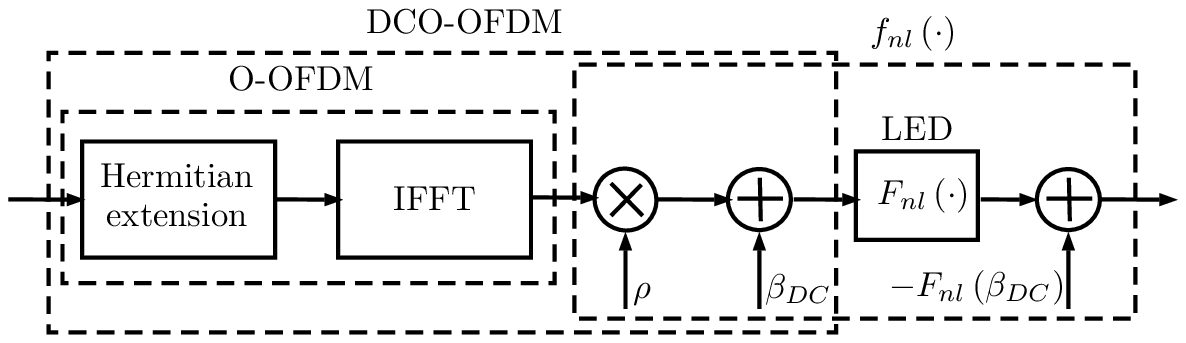}
\caption{\label{fig:OFDMDiagram} Block diagram showing how the physical model is handled for analysis.}
\end{figure}
\fi

Since we are not making considerations about the amount of DC added power or the joint usage of the system for lighting and information transfer, for fairness we will assume in the sequel a biasing value $\beta_{DC}$ placing the zero-mean signal $\mathrm{X}_i$ in the middle of the transfer function dynamic range. This is the best situation respecting the effects from lower and upper clipping. This is what would be done when lighting applications are not considered. Note that, in this case, both $\mathrm{X}_i$ and $\mathrm{Z}_i$ are zero-mean sequences.

The transmitted sequence $\mathrm{Z}_i$ goes through an additive white Gaussian noise (AWGN) channel with power spectral density $N_0/2$, and the signal at the receiver input is
\begin{equation}
\mathrm{R}_i=\mathrm{Z}_i + n_i,
\end{equation}
where $n_i$ is the zero-mean noise sample. According to the context chosen, there are no mobility effects or interferences considered in the channel model. We do not include a model for the photodiode at the receiver, since its response can be considered fully linear: therefore, the only optical channel effects considered here are the ones corresponding to the transmitter LED \cite{elgala2009non}. The received sequence $\mathrm{R}_i$ is O-OFDM demodulated and deinterleaved to get the sequence $r_i$, which, in turn, is processed by the chaos-based coded demodulator (where knowledge of the conjugation function used in the transmitter is the only requisite). The result is the estimated bit sequence $b'_i$.

\section{Error probability analysis}
\label{analysis}

Given the nature of the CCM encoding and decoding, error bounds can be developed for a variety of channels by just taking into account their characteristic error loops. This feature has been extensively exploited in AWGN and fading channels \cite{Escribano09,Escribano14}. More recently, the ability to calculate a tight bound based on the lower length error loops has been used to create a cost function apt to optimize the BER response in presence of an HPA nonlinearity in a standard RF system \cite{Escribano16b}. The basis for this development has been the definition and optimization of a conjugation function $g\left(\cdot\right)$ that suitably modifies the waveform shaping of the output CCM samples $x_i$. Though the approach is based on the same principles, there are important differences with respect to the mentioned work in \cite{Escribano16b}: in this case, knowledge about the exact waveform received for each error loop was necessary in order to characterize an equivalent Euclidean distance. As we are going to explain in the sequel, for the current nonlinear LED setup with OFDM, there is no need to have the received waveform, only the characterization of the nonlinear LED response, and other basic parameters of the communication. Therefore, the resulting algorithm has important differences induced by the differences in the context and application.

The presence of the O-OFDM modulator and demodulator, along with the limited range of the LED response, produces a thorough modification of the statistics of the signals involved \cite{Elgala10}. The signal $x_i$ has power $\mathrm{E} \left[ x_i^2 \right]=\sigma_x^2$, and, if $N$ is high enough, so does $X_i$, $\mathrm{E} \left[ \mathrm{X}_i^2 \right] \simeq \sigma_x^2$. This is due to the fact that the effect of the two subcarriers bearing nulls after the Hermitian extension, those located at positions $0$ and $N/2$, can be considered negligible in practice.

By virtue of the central limit theorem and the uncorrelation of the samples entering the O-OFDM modulator\footnote{This uncorrelation is guaranteed by the interleaver.}, $\mathrm{X}_i$ turns out to be a zero-mean Gaussian distributed random variable, and $\mathrm{Z}_i$ can be characterized, with the help of the Bussgang theorem, as \cite{Dimitrov13}
\begin{equation}
\mathrm{Z}_i = C \cdot \mathrm{X}_i + \eta_i,
\end{equation}
where $C$ is a real-valued constant and $\eta_i$ is a zero-mean non-Gaussian noise sequence. Therefore, the effect of the LED nonlinearity amounts to linearly modifying the amplitude of the information bearing signal\footnote{Note that $C$ also accounts for the effect of the back-off applied.}, and the addition of a disturbance with power $\mathrm{E}\left[ \eta_i^2 \right]=\sigma_{\eta}^2$. The coefficient $C$ can be calculated as
\begin{equation}
\label{eq1}
C=\frac{\mathrm{E}\left[ \mathrm{X}_i \cdot \mathrm{Z}_i \right]}{\sigma_x^2},
\end{equation}
and the variance of $\eta_i$ as
\begin{equation}
\label{eq2}
\sigma_\eta^2=\mathrm{E}\left[ \mathrm{Z}_i^2 \right]-C^2 \cdot \sigma_x^2.
\end{equation}
Under these hypothesis, the signal at the input of the receiver is
\begin{equation}
\mathrm{R}_i=C \cdot \mathrm{X}_i + \eta_i + n_i.
\end{equation}
After going through the O-OFDM demodulator and the deinterleaver, the recovered sequence can be written as
\begin{equation}
\label{eqr}
r_i=C \cdot x_i + \xi_i + \chi_i,
\end{equation}
where $x_i$ is affected by the same factor $C$ as $\mathrm{X}_i$, and $\xi_i$ and $\chi_i$ are the FFT filtered and deinterleaved versions of $\eta_i$ and $n_i$. Given the central limit theorem and the nature of the O-OFDM demodulation, both $\xi_i$ and $\chi_i$ turn out to be independent samples of zero-mean AWGN processes, with variances
\begin{eqnarray}
\mathrm{E}\left[ \xi_i^2 \right] = \sigma_{\eta}^2, \\
\mathrm{E}\left[ \chi_i^2 \right] = \sigma_n^2.
\end{eqnarray}
Note that this result is true regardless where the biasing and de-biasing operations take place, while the analysis itself can only be carried out with the help of the Bussgang theorem if $f_{nl}\left(\cdot\right)$ is as previously defined, without loss of generality.

If the samples at the output of the receiver follow \eqref{eqr}, where the perturbations are additive, Gaussian and uncorrelated, we can draw a bound based on the pairwise error probability (PEP) depending on the most probable error loops. If we send the information $\mathbf{b}=\left(b_1,\cdots,b_M\right)$, corresponding to chaotic sequence $\mathbf{x}=\left(x_1,\cdots,x_M\right)$, the decoder will fail according to a given error loop $\mathbf{e}=\left(0, \cdots, 0, e_j, \cdots, e_{j+L_{\mathbf{e}}-1}, 0, \cdots, 0 \right)$ of length $L_{\mathbf{e}}$, when the erroneous chaotic sequence $\mathbf{x}'=\left(x'_1,\cdots,x'_M\right)$ corresponding to the erroneous decoded information $\mathbf{b}'=\left(b'_1,\cdots,b'_M\right)=\mathbf{b}+\mathbf{e}$ meets
\begin{equation}
\sum_{i=1}^M | r_i - C \cdot x'_i |^2 < \sum_{i=1}^M | r_i - C \cdot x_i |^2,
\end{equation}
where we assume the availability at the receiver of an accurate estimation of the coefficient $C$\footnote{This channel state information (CSI) scenario is usual in OFDM, and it is frequently performed with the help of pilots and training sequences.}. After some algebra, it can be demonstrated that this allows the calculation of the PEP as
\begin{equation}
\label{eqPEP}
P\left(\mathbf{x} \rightarrow \mathbf{x}' | \mathbf{x}, \mathbf{e} \right) = \frac{1}{2} \mathrm{erfc} \left( \frac{C \cdot d_E\left(\mathbf{x},\mathbf{x}'\right)}{2\sqrt{2 \left( \sigma_{\eta}^2+\sigma_n^2 \right)}} \right),
\end{equation}
where
\begin{equation}
\label{eucdist}
d_E\left(\mathbf{x},\mathbf{x}'\right)=\sqrt{\sum_{i=1}^M |x_i-x'_i |^2},
\end{equation}
is the Euclidean distance between sequences $\mathbf{x}$ and $\mathbf{x}'$. This PEP may provide a good approximation when SNR is high enough, so that the actual MAP decoding performed for the CCM \cite{Escribano06b} tends to the optimal ML decoding.

By considering a set $B_{\mathbf{e}}$ of most probable error loops, a bound on the error probability can be thus calculated under the theory of rare events as
\begin{equation}
\label{eq3}
P_b \leq \widehat{P}_b \triangleq \sum_{\mathbf{e} \in B_{\mathbf{e}}} \sum_{\mathbf{x}} \frac{\omega\left( \mathbf{e} \right)}{2^{Q+L_{\mathbf{e}}}} P\left(\mathbf{x} \rightarrow \mathbf{x}' | \mathbf{x}, \mathbf{e} \right),
\end{equation}
where $\omega\left( \mathbf{e} \right)$ is the binary weight of the error loop. Data vector $\mathbf{x}=\left(x_1,\cdots,x_{L_{\mathbf{e}}} \right)$ only considers the length of the loop, and there are up to $2^{Q+L_{\mathbf{e}}}$ different equiprobable data sequences. Due to the fact that CCM does not satisfy the uniform error property (UEP) of other TCM systems, it is not possible to simplify the bound with the help of the usual minimum distance analysis. Therefore, this bound has to be calculated numerically. This means that the number and length of the loops considered to perform the averaging expressed in equation \eqref{eq3} has to be limited for computational convenience.

On the other hand, as the power of the part of the signal actually bearing the information is $C\cdot \sigma_x^2$, the AWGN noise power in the channel can be related to the $E_b/N_0$ parameter as
\begin{equation}
\sigma_n^2=\frac{C^2 \cdot \sigma_x^2}{2\left.\frac{E_b}{N_0}\right|_{\mathrm{AWGN}}},
\end{equation}
where we have only taken into account the distortion given by the Gaussian noise, as indicated by the AWGN label. For convenience, we do not include the biasing power in the calculations, since this value will be fixed for all the scenarios tested, so that the comparisons will be fair and will not lose generality. On the other hand, we may define an equivalent received per bit signal-to-noise ratio by taking into account as well the nonlinear distortion equivalent noise $\sigma_{\eta}^2$, as
\begin{equation}
 \left.\frac{E_b}{N_0}\right|_{eq}=\frac{C^2 \cdot \sigma_x^2}{2\left(\sigma_{\eta}^2 + \sigma_n^2 \right)}.
\end{equation}
Such parameter would be the one actually measured at the receiver with the help of the average error vector magnitude (EVM) parameter, and would have an impact in measuring the behavior of the system in a practical setup.

In the sequel, we are going to analyze how to calculate expressions \eqref{eq1} and \eqref{eq2}. Given that $\mathrm{X}_i$ follows a zero-mean Gaussian probability density function (pdf) with variance $\sigma_x^2$, we have
\begin{equation}
\mathrm{E}\left[ \mathrm{X}_i \cdot \mathrm{Z}_i \right] = \frac{1}{\sqrt{2\pi\sigma_x^2}} \int_{-\infty}^{\infty} x f_{nl}\left( x \right) \mathrm{e}^{-\frac{x^2}{2\sigma_x^2}} dx.
\end{equation}
Given the definition of $f_{nl}(x)$, \eqref{eq0}, when $F_{nl}\left( \cdot \right)$ takes the form of a symmetrical polynomial around the center of the dynamic range with degree $n$, $f_{nl}\left( \cdot \right)$ turns out to be another related polynomial function of degree $n$
\begin{equation}
f_{nl}\left(x\right)=\sum_{l=0}^{n} a_l x^l,
\end{equation}
symmetric around the origin, and with minimum and maximum values (lower and upper clipping levels)
\begin{eqnarray}
&f_{nl}\left(\lambda_D \right)=-F_{nl}\left( \beta_{DC} \right),& \\
&f_{nl}\left(\lambda_U \right)=F_{nl}\left( \beta_{DC} \right),&
\end{eqnarray}
respectively. Values $\lambda_U=-\lambda_D$ are the input thresholds for the LED nonlinear response between the clipping levels. After some algebra, it is easy to demonstrate \cite{Dimitrov13}
\begin{equation}
C=\frac{1}{\sigma_x^2} \left( 2 f_{nl}\left( \lambda_U \right) \sigma_x \frac{\mathrm{e}^{-\frac{\lambda_U^2}{2\sigma_x^2}}}{\sqrt{2\pi}} + \sum_{l=0}^{n} a_l I_{l+1} \right),
\end{equation}
where
\begin{equation}
I_j=\left\{ \begin{array}{ll} 0 & \mathrm{odd} \ j \\ -\frac{2 \lambda_U^{j-1}\sigma_x}{\sqrt{2\pi}} \mathrm{e}^{-\frac{\lambda_U^2}{2 \sigma_x^2}}+\left( j-1 \right) \sigma_x^2 I_{j-2}, & \mathrm{even} \ j \end{array} \right. ,
\end{equation}
and
\begin{equation}
I_0=1-\mathrm{erfc}\left( \frac{\lambda_U}{\sqrt{2}\sigma_x} \right).
\end{equation}
On the other hand, $\mathrm{E}\left[ \mathrm{Z}_i^2 \right]$ can be calculated after a similar algebraic procedure as \cite{Dimitrov13}
\begin{equation}
\mathrm{E}\left[ \mathrm{Z}_i^2 \right] = f_{nl}^2\left( \lambda_U \right) \mathrm{erfc} \left( \frac{\lambda_U}{\sqrt{2}\sigma_x} \right) + \sum_{l=0}^{n} \sum_{k=0}^{n} a_l a_k I_{l+k}.
\end{equation}
With this value and $C$, we can obtain \eqref{eq2} and, finally, calculate \eqref{eq3}.

Note that the results obtained with this framework would be progressively better for increasing $N$ (FFT order in the O-OFDM), when the assumption of gaussianity based on the central limit theorem gets tighter. At this point we have all the tools needed to calculate a bound on the bit error probability for the setup presented in Figure \ref{fig:BlockDiagram}, as a function of a given CCM, a given conjugation function $g\left( \cdot \right)$, and a given IBO. In the sequel, we are going to explain how to parametrize $g\left(\cdot\right)$, and how to address the optimization process in order to counteract the effects of the LED nonlinearity on the system BER.

\section{Optimization}
\label{optim}

In order to design the CCM system, we propose the optimization of the conjugation function $g\left(\cdot\right)$ with the objective of improving the system binary error probability (BEP), including AWGN and the nonlinear effects of the LED. We approximate the BEP by means of the bound $\widehat{P}_b$ \eqref{eq3}, which depends on the conjugation function through \eqref{eq:fconj}, for a fixed number of possible error loops up to a given length $\left.L_{\mathbf{e}}\right|_{\max}$. In order to provide a numerical approximation of $g\left(\cdot\right)$, we sample its domain $[0,1]$ with $P$ equidistant samples. The final conjugation function may be thus obtained by interpolation techniques. The samples are $z^j=j/P, ~ j=0,\ldots,P$, matched to the function samples $s^j=g\left(z^j\right)$. We adopt this notation to distinguish these variables from the related temporal sequences $z_i$ and $s_i$. Due to the nature of the conjugation function $g$, the discretized values $s^j$ and $z^j$ must meet the constraints described in the system setup. The formulation of the optimization problem is given by
\begin{equation}
\label{eq:optimization}
\begin{array}{clr}
\displaystyle \min_{s^j} & \lbrace \widehat{P}_b \rbrace  & \\
\textrm{s.t.} 	& 0  <  z^j  <  1, 	\\
& 0  <  s^j  <  1, 	\\
& s^j < s^{j+1},	\\
\end{array}
\end{equation}
where $j=1,\cdots,P-1$.

This problem consists in a classical minimization of a nonlinear objetive function with linear constraints. We apply the Interior Point Algorithm implemented in MATLAB Optimization Toolbox, described in \cite{byrd2000} and \cite{waltz2006}. After a number of tests using random starting points, the optimization converges always to the same solution. For practical reasons, the seed used is a linear function, i.e. the sampling of $g\left(z\right)=z$. Note that the optimization is performed for the system design, so the processing time is not a problem when operating the communication system after the initial setup.

\section{Results and discussion}
\label{results}
As detailed in the system description, the specific CCM system chosen for illustration of the proposal is the mTM CCM. The parameters chosen have been $Q=6$ for the quantization parameter, and the error loops considered have been all possible ones up to maximum length $\left.L_{\mathbf{e}}\right|_{\max}=2 Q$ (which, after analysis, makes a total of $32$ error loops). The number of points considered in the discretization of $g\left(\cdot\right)$ has been $P=64$, and we have set a target $\left.E_b/N_0\right|_{\mathrm{AWGN}}$ value of $10$ dB\footnote{Through previous simulation work (not shown here for brevity), it has been verified that the SNR value does not affect the optimization result. The specific SNR chosen is appropriate from the point of the view of the corresponding PEP values in (\ref{eqPEP}), which should not be too small nor too high, since this could lead to numerical problems.} in the optimization of bound \eqref{eq3}. Values of the quantization parameter from $4$ to $6$ have shown to be appropriate to keep the chaotic characteristics of the system, while avoiding excessive complexity at the encoding and decoding processes without affecting final performance (which should be independent from $Q$) \cite{Escribano09}. The O-OFDM framework uses an FFT of order $N=256$, and the input data block length is $M=12700$, equivalent to $100$ O-OFDM symbols.

In order to model the LED response, we choose for $F_{nl}\left( \cdot \right)$ a known polynomic transfer function, namely the one considered in \cite{Dimitrov13},
\begin{equation}
\label{eq00}
F_{nl}\left( x \right) = \left\{ \begin{array}{ll} 0.0, & x \leq 0.1 \\ -1.6461 x^3 + 2.7160 x^2 - & \\ -0.4938 x + 0.0239, & 0.1 < x \leq 1.0 \\ 0.6, & x > 1.0 \end{array} \right.
\end{equation}
The bias level considered will thus be $\beta_{DC}=0.55$ in all the cases. We use this LED response to perform the system design, the  performance simulations and the comparisons with classical alternatives.

Taken into account the described CCM system and its parameters, the optimization problem of equation \eqref{eq:optimization} and the LED response \eqref{eq00}, the resulting optimized conjugation function is given in Figure \ref{Sim_fig_1}. We can see the conjugation functions obtained for IBO=$0$, $10$ and $40$ dB, meaning high, moderate and low nonlinearity, respectively. The staircase form for all the cases means that the constellation values for the phase mapped mTM will be located around four positions, with slight changes in the level of some of their specific values as a function of the IBO factor. This shows the consistency of the optimized solution.
\begin{figure}[htb!]
\centering
\ifonecol
\includegraphics[width=0.75\textwidth, keepaspectratio]{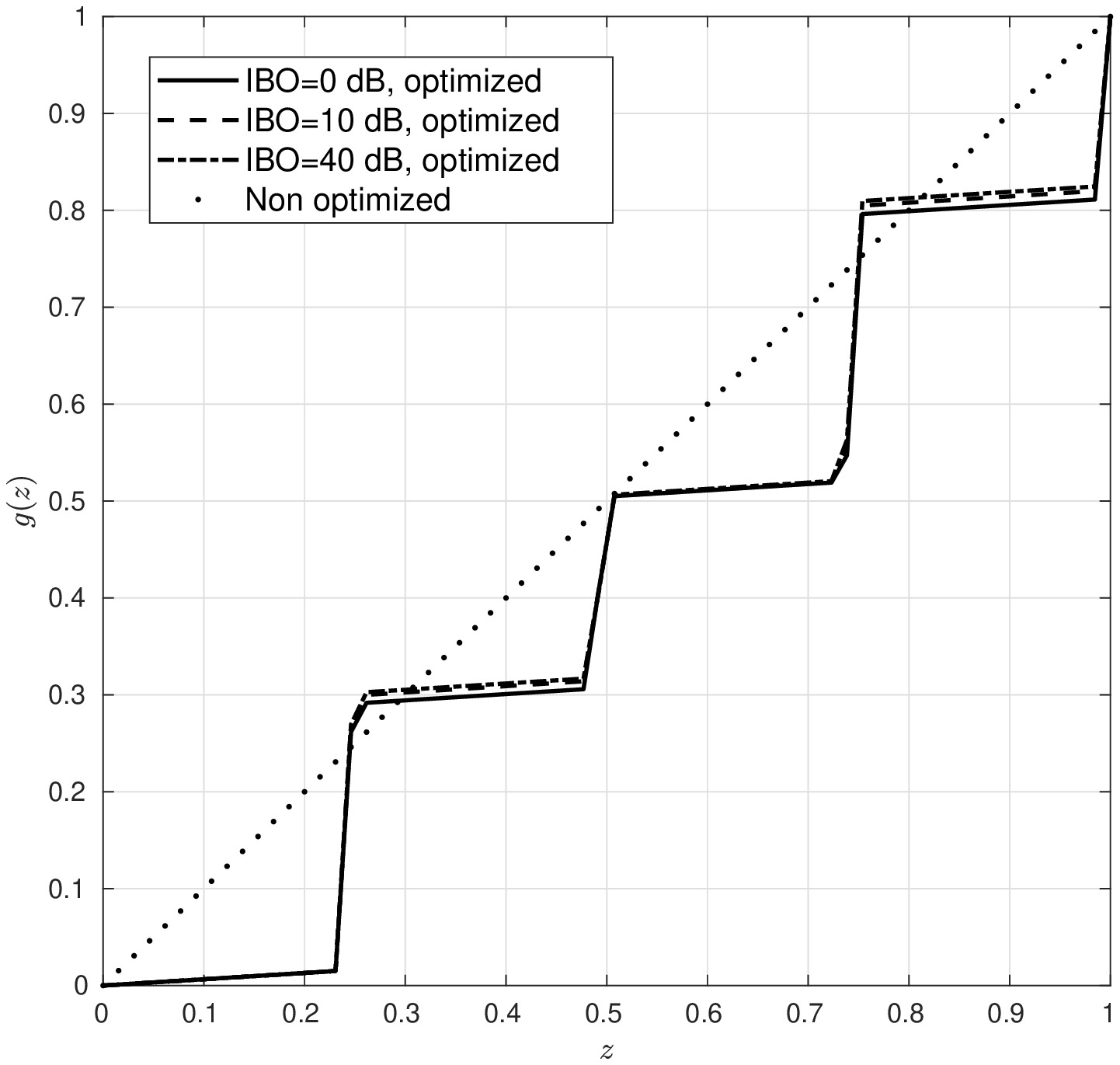}
\caption{Optimized conjugation functions for several IBO cases, when using the mTM CCM with $Q=6$, $\left.L_{\mathbf{e}}\right|_{\max}=2 Q$, $P=64$, $E_b/N_0=10$ dB.
\label{Sim_fig_1}}
\end{figure}
\else
\includegraphics[width=\figurewidth, keepaspectratio]{18-0201_fig3.eps}
\caption{Optimized conjugation functions for several IBO cases, when using the mTM CCM with $Q=6$, $\left.L_{\mathbf{e}}\right|_{\max}=2 Q$, $P=64$, $E_b/N_0=10$ dB.
\label{Sim_fig_1}}
\end{figure}
\fi
On the other hand, in Figure \ref{Sim_fig_2} we show the normalized histogram of the resulting squared Euclidean distances, $d_E^2\left(\mathbf{x},\mathbf{x}'\right)$, for the cases considered in Figure \ref{Sim_fig_1}. Without optimization (when the conjugation function is just the identity), we have a Gaussian-like shape for the distance spectrum, excepting some discrete values concentrated in the lower part, amounting for the minimum distance that would determine the BER behavior for high SNR. In the optimized cases, the shape of the distance spectrum changes dramatically, but keeping the same form in all the cases, as may be expected for the homogeneity of the resulting conjugation functions. It is to be noted that, in these cases, the minimum distance is higher, and there is an irregular concentration of values around some specific distance points, with maxima in the middle and the upper part of the distribution. This explains the improvements in performance that we will comment in the sequel. Results after sweeping other values for parameters $Q$, $\left.L_{\mathbf{e}}\right|_{\max}$, $P$, $E_b/N_0$ and $N$ (not shown), demonstrate the consistency of the approach, because no improvement is obtained with respect to the parameter set represented in the plot.
\begin{figure}[htb!]
\centering
\ifonecol
\includegraphics[width=0.75\textwidth, keepaspectratio]{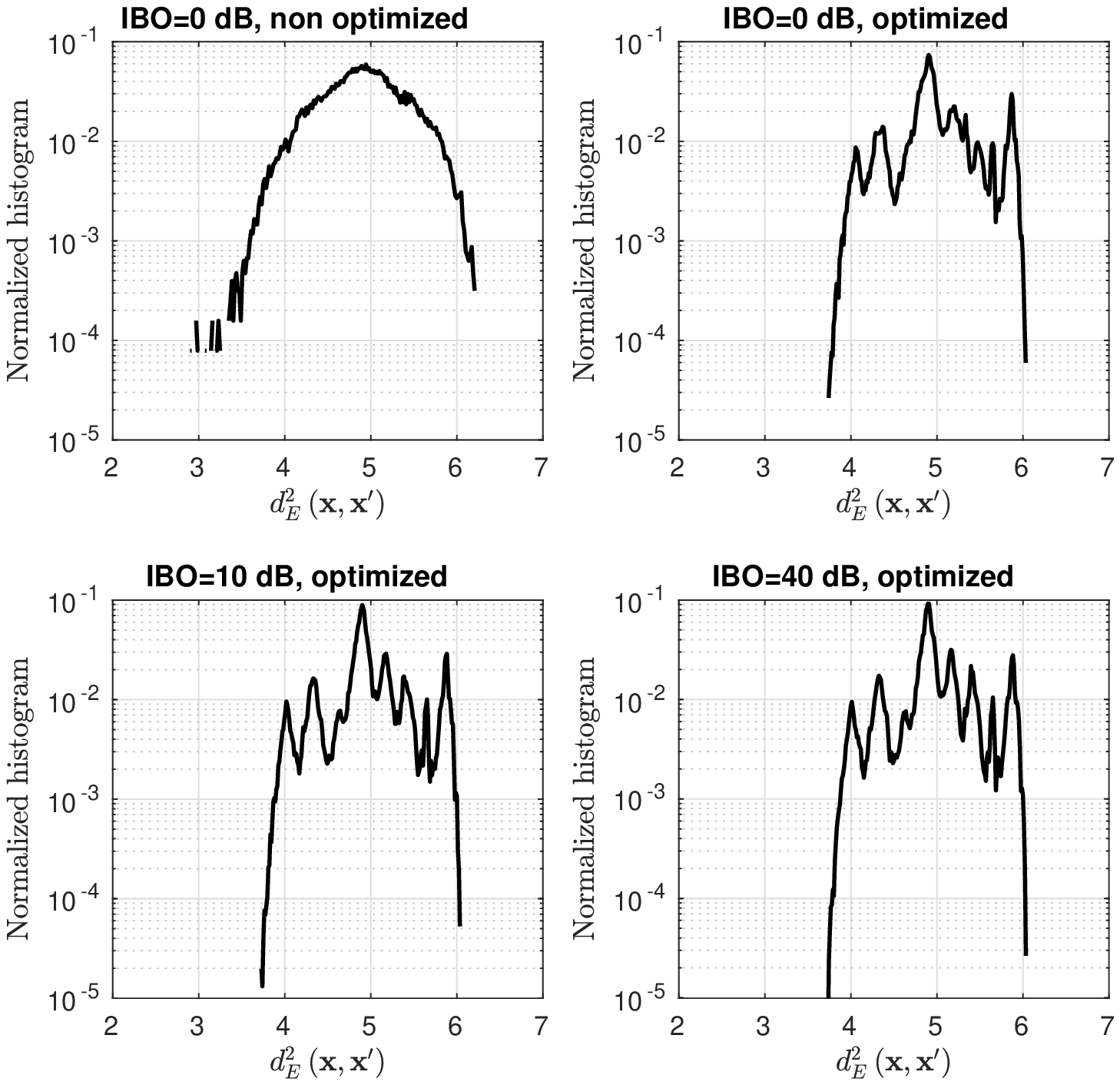}
\caption{Normalized histograms of squared Euclidean distances $d_E^2\left(\mathbf{x},\mathbf{x}'\right)$ for the mTM CCM with the chosen set of parameters.
        \label{Sim_fig_2}}
\end{figure}
\else
\includegraphics[width=\figurewidth, keepaspectratio]{18-0201_fig4.eps}
\caption{Normalized histograms of squared Euclidean distances $d_E^2\left(\mathbf{x},\mathbf{x}'\right)$ for the mTM CCM with the chosen set of parameters.
	\label{Sim_fig_2}}
\end{figure}
\fi

On the other hand, to show the power of the present approach, we are also going to consider the classical alternatives under ideal predistortion, where the linearized resulting LED transfer function will be
\begin{equation}
\label{eq01}
F_{pred}\left( x \right) = \left\{ \begin{array}{ll} 0.0, & x \leq 0.1 \\ 0.75 x - 0.075, & 0.1 < x \leq 1.0 \\ 0.6, & x > 1.0 \end{array} \right. .
\end{equation}
The bias level required will be the same. Note that in the ideally predistorted response, there is only the clipping as distorting factor. As stated in the introductory Section, a CCM is a kind of nonlinear non-uniform TCM-like system, and therefore TCM is the natural classical counterpart to be compared with. In our case, we have chosen a $16$-QAM scheme, since high efficiency modulations are preferred for multicarrier schemes. However, as the overall efficiency of the CCM system is $1$ bit per symbol, for fairness we consider here a rate-$1$ TCM scheme, built by means of a rate-$1/4$ nonrecursive nonsystematic convolutional code (CC) with polynomials
\begin{eqnarray}
\mathbf{g}_1=0127,\ \mathbf{g}_2=0171,\ \mathbf{g}_3=0155,\ \mathbf{g}_4=0177,
\end{eqnarray}
in octal form. Therefore, this represents the most favorable scenario for a TCM of this kind, from the point of view of final performance. Complexities of the systems compared are the same, since both consist in cascading a $64$-state binary trellis encoder and a mapping operation, that require the same consumption of resources in transmission and reception.

\begin{figure}[htb!]
\centering
\ifonecol
\includegraphics[width=0.75\textwidth, keepaspectratio]{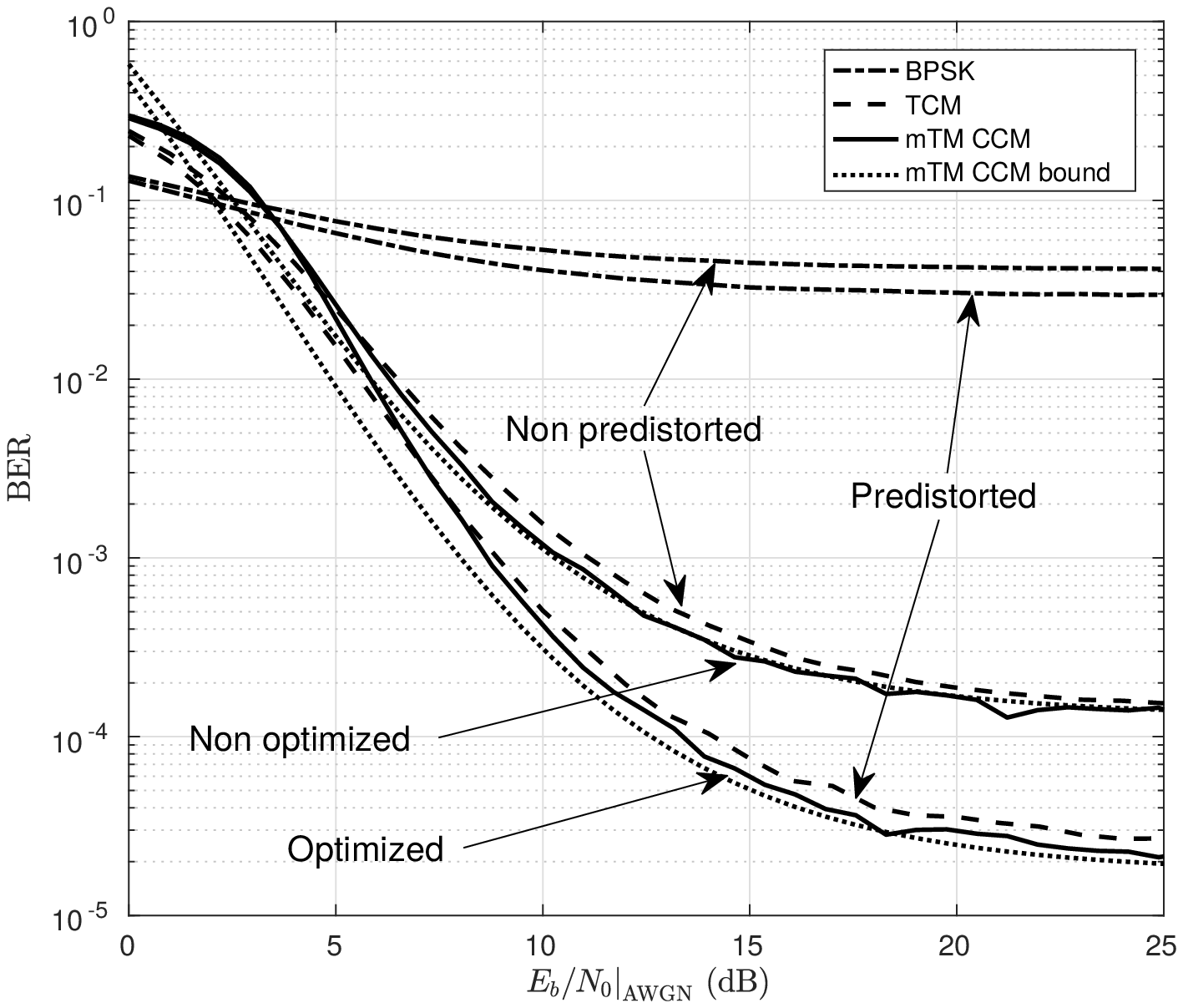}
\caption{Results for IBO=$0$ dB. Dash-dotted lines: uncoded BPSK. Dashed lines: TCM. Continuous lines: mTM CCM. Dotted lines: bounds for mTM CCM.
\label{Sim_fig_3}}
\end{figure}
\else
\includegraphics[width=\figurewidth, keepaspectratio]{18-0201_fig5.eps}
\caption{Results for IBO=$0$ dB. Dash-dotted lines: uncoded BPSK. Dashed lines: TCM. Continuous lines: mTM CCM. Dotted lines: bounds for mTM CCM.
\label{Sim_fig_3}}
\end{figure}
\fi

\begin{figure}[htb!]
\centering
\ifonecol
\includegraphics[width=0.75\textwidth, keepaspectratio]{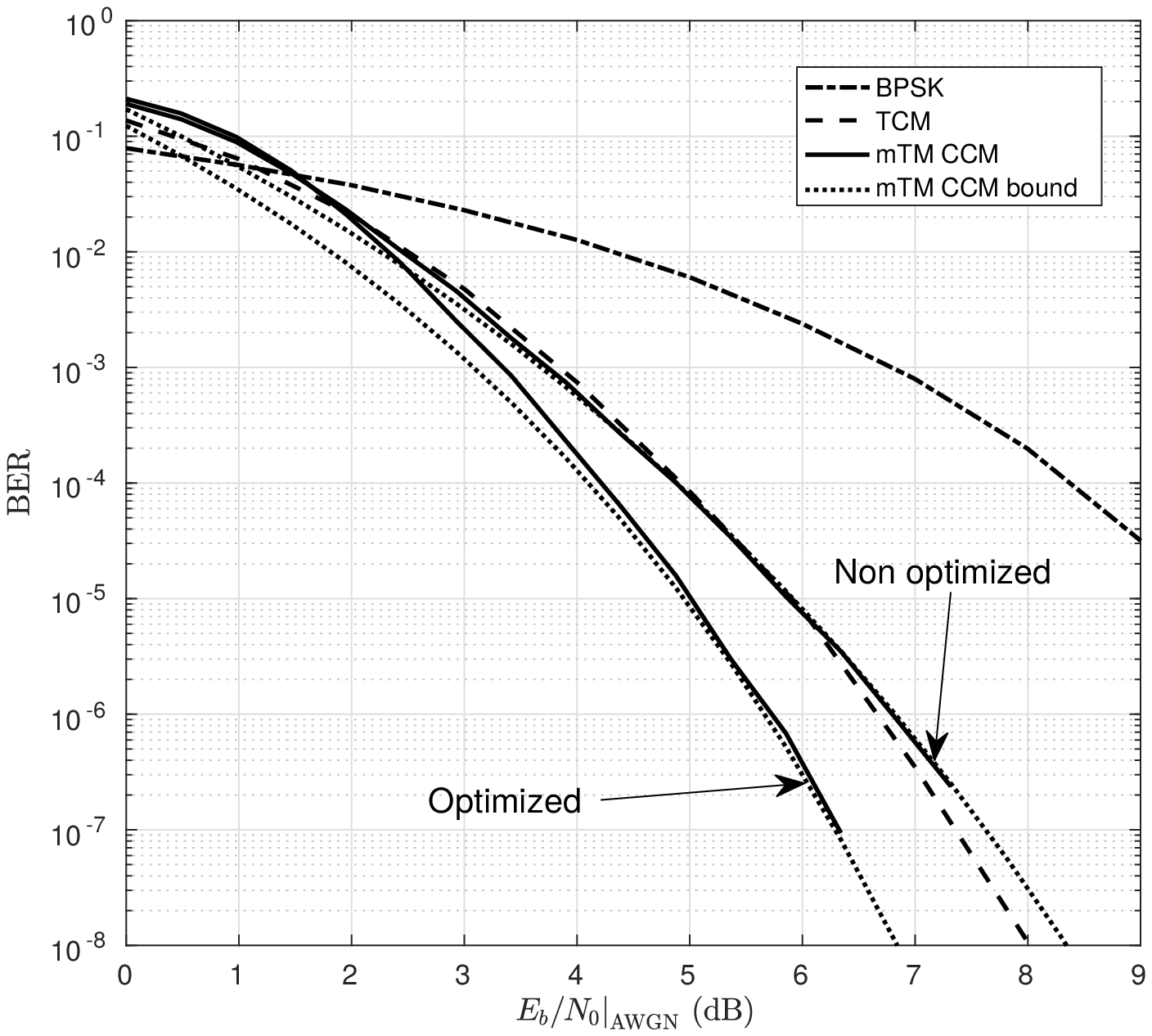}
\caption{Results for IBO=$40$ dB. Dash-dotted line: uncoded BPSK. Dashed line: TCM. Continuous lines: mTM CCM. Dotted lines: bounds for mTM CCM.
\label{Sim_fig_4}}
\end{figure}
\else
\includegraphics[width=\figurewidth, keepaspectratio]{18-0201_fig6.eps}
\caption{Results for IBO=$40$ dB. Dash-dotted line: uncoded BPSK. Dashed line: TCM. Continuous lines: mTM CCM. Dotted lines: bounds for mTM CCM.
\label{Sim_fig_4}}
\end{figure}
\fi
In Figure \ref{Sim_fig_3}, we have depicted the BER results and the bounds for several cases, when IBO=$0$ dB. The classical alternatives (BPSK, TCM) have been simulated with and without ideal predistortion. Uncoded BPSK offers a very poor performance in both cases, as may be expected. The TCM scheme gets a steady improvement when considering ideal predistortion. In the case of the mTM CCM, the non-optimized setup offers a performance slightly better than the TCM scheme, and this advatange is kept after optimization. It is to be noted that the error floors reached have to do with the equivalent noise power $\sigma_{\eta}^2$ determined by the LED nonlinearity, which is a constant value. Very importantly, the plots show the excellent agreement between mTM CCM simulation results and bounds (specially from $10$ dB $E_b/N_0$ and on), thus validating the usefulness and coherence of the approach. In Figure \ref{Sim_fig_4}, we have represented the same plots when IBO=$40$ dB (thus providing information about the extreme case when the nonlinearity is negligible). In the case of the classical alternatives, there is only one curve, since the ideally predistorted and non-predistorted cases converge. Uncoded BPSK is again non competitive, whereas the TCM results are very close to the mTM CCM results without optimization. After optimization, the chaos-based system clearly outperforms TCM, demonstrating that the approach can be also useful under just AWGN. Again, the bounds are very tight, in this case from $4$ dB $E_b/N_0$ and on.

\begin{figure}[htb!]
\centering
\ifonecol
\includegraphics[width=0.75\textwidth, keepaspectratio]{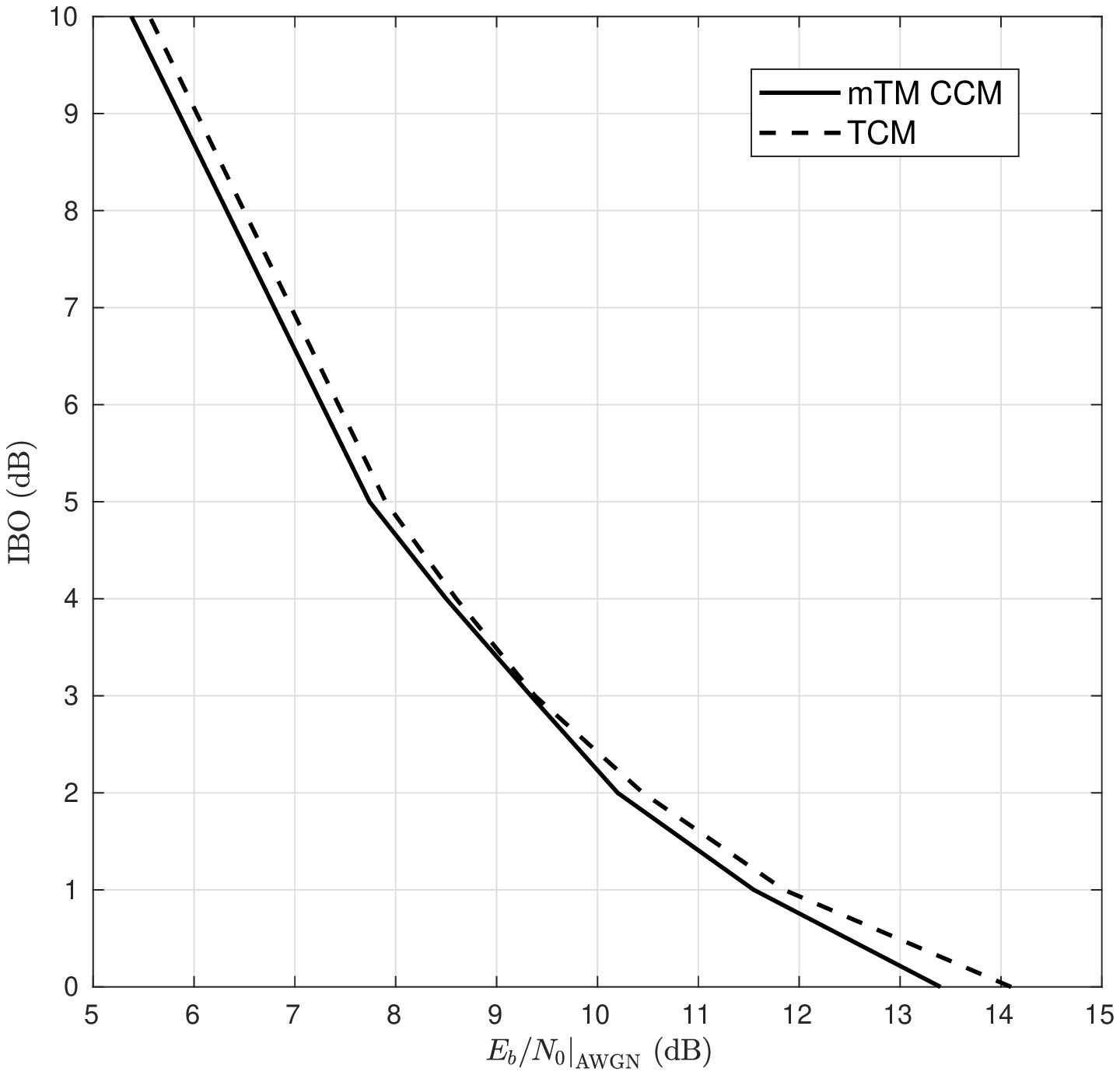}
\caption{Comparative of IBO and $E_b/N_0$ required for a BER of $10^{-4}$: optimized mTM CCM against ideally predistorted TCM.
\label{Sim_fig_5}}
\end{figure}
\else
\includegraphics[width=\figurewidth, keepaspectratio]{18-0201_fig7.eps}
\caption{Comparative of IBO and $E_b/N_0$ required for a BER of $10^{-4}$: optimized mTM CCM against ideally predistorted TCM.
\label{Sim_fig_5}}
\end{figure}
\fi
To get a better idea of the evolution of the BER results as a function of IBO and $E_b/N_0$, we have considered a target BER value of $10^{-4}$, and we have plotted the results for the optimized mTM CCM against the ideally predistorted TCM setup, as may be seen in Figure \ref{Sim_fig_5}. We can verify that there is a slight gain for the chaos-based system along all the IBO factors considered, reaching a minimum value in the middle range, and becoming higher on the extremes. The gain for IBO=$0$ dB (worst case considered) is around $0.7$ dB in $E_b/N_0$. For IBO=$40$ dB (AWGN case), the gain is almost $1$ dB (point not shown). It is worth noting that the best possible predistortion case has been considered, a strictly linear effective response. A real predistortion system includes discretization errors that may reduce its performance and improve the relative comparison with the proposed system.

\section{Practical implementation aspects}
\label{implementation}

Here we give a glimpse on the practical implementation of the method proposed. On boot-up, the nonlinearity of the LED can be estimated offline using already known methods \cite{Dong16}. The characteristic obtained can easily and automatically be fitted using a polynomial model, and then the optimization step can be performed as previously described in order to get the suitable conjugation function. Said function can be stored in a look-up table (LUT) at the transmitter, and the corresponding data can be sent to the receiver using a reliable side channel. An instance of this side channel could be implemented by using a single-carrier robust low rate modulation (BPSK, for example), as a part of the initial protocol to establish the communication. This is a known and very usual strategy to handle access in communications.

Once the receiver has the information about the conjugation function, the chaos-based O-OFDM system can work as explained. An additional feature that could be considered to make the system more robust consists in storing a number of possible conjugation functions in LUT form, matched to several possible working conditions of the LED. Moreover, we can think of tackling the problem of drifting due to heating or aging, which may be done by means of a periodic assessment of the LED condition through system identification (as in the boot-up process described), and a subsequent re-adaptation of the conjugation function LUT. Note that the complexity of all these strategies lies in the initial protocol itself, as the usage of a LUT requires very limited resources consumption.

\section{Conclusions}
\label{conclusions}
In this article, we have developed a chaos-based communication framework adapted to efficiently compensate the nonlinear distortion of LED OWC (or VLC) transmission. The approach is similar to predistortion, but it offers significant differences. In predistortion, a given waveform parameter is taken heed of (like PAPR or some distortion measurement), whereas in our approach it is the BER itself which is optimized through a process that encompasses the whole communication chain, not just the transmitter. As a result of the characterization of the system, a practical bound has been developed, which is very tight when compared with the simulation results. Based on this semi-analytic bound, the optimization of a parametrized conjugation function applied to the CCM has led to a practical design criterion. The results, compared to ideal predistortion and classical alternatives, show that the final BER can be advantageously minimized. The process described here can be applied prior to any transmission, through identification of the specific LED setup transfer function. Future work will focus on extending this strategy to other kinds of OWC multicarrier frameworks, where the same problems hinder the efficiency of the system.


%

%

%

\ifCLASSOPTIONcaptionsoff
  \newpage
\fi

\begin{IEEEbiography}[{\includegraphics[width=1in,height=1.5in,clip,keepaspectratio]{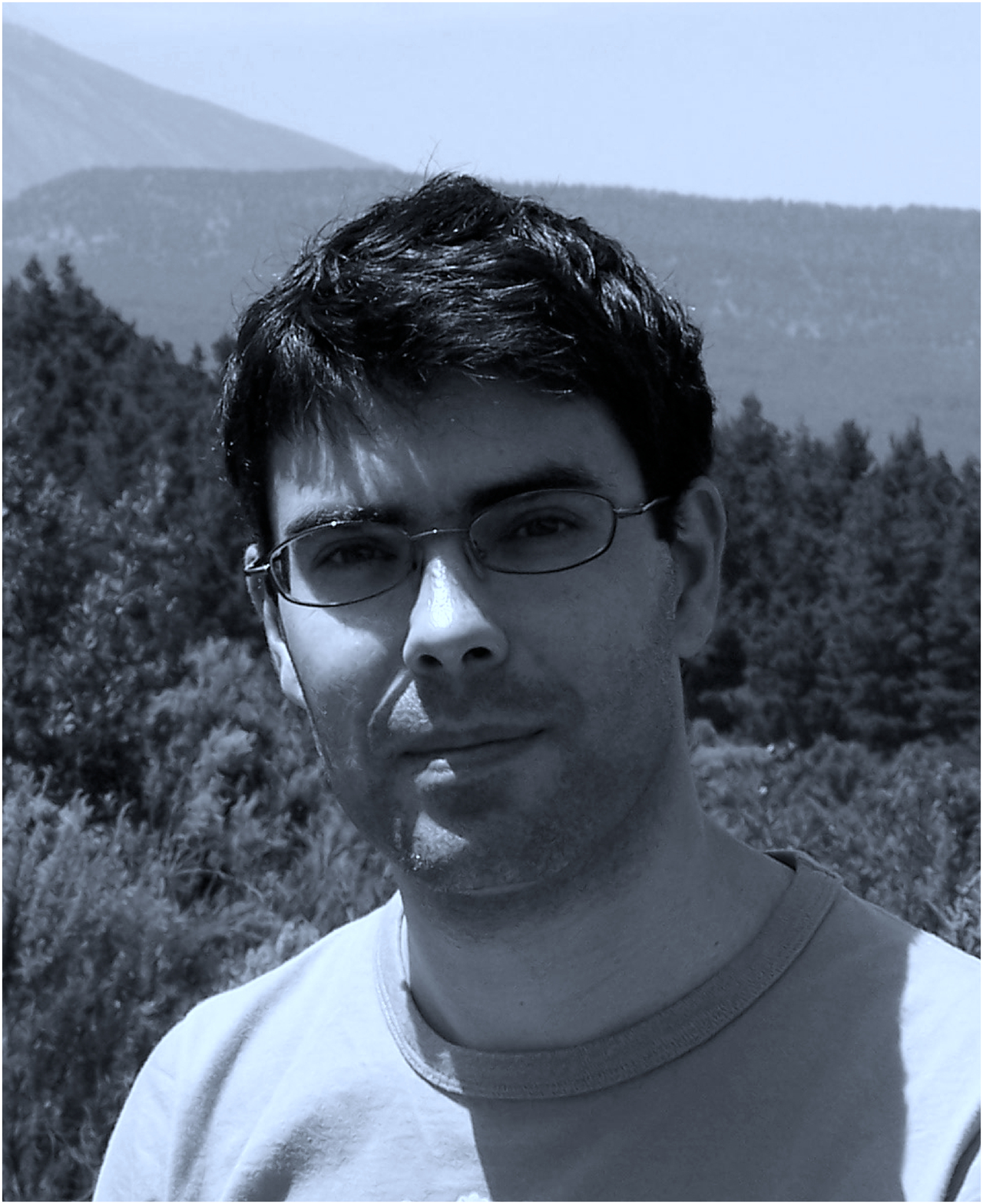}}]
{Francisco J. Escribano} (M'06, SM'16) received his degree in Telecommunications Engineering at ETSIT-UPM, Spain, and his PhD degree at Universidad Rey Juan Carlos, Spain. He is currently Associate Professor at the Department of Signal Theory and Communications of Universidad de Alcal\'{a}, Spain, where he is involved in several undergraduate and master courses in Telecommunications Engineering. He has been Visiting Researcher at the Politectnico di Torino, Italy, and at the EPFL, Switzerland. His research activities are focused on Communications Systems and Information Theory, mainly on the topics of channel coding, modulation and multiple access, and on the applications of Chaos in Engineering.
\end{IEEEbiography}

\begin{IEEEbiographynophoto}
{Jos\'e S\'aez-Landete} was born in Valdeganga (Albacete), Spain, in 1977. He received the M.S. degree in Physics from the Universidad de Zaragoza, Spain, in 2000 and the Ph.D. degree in Physics from the Universidad Complutense de Madrid, Spain, in 2006. Since 2006, he has been with the Signal Theory and Communications of the Universidad de Alcal\'a, Madrid, Spain, where he is an Associate Professor. His research interests include digital signal processing, image processing, filter design, digital communications and optimization.
\end{IEEEbiographynophoto}

\begin{IEEEbiography}[{\includegraphics[width=1in,height=1.25in,clip,keepaspectratio]{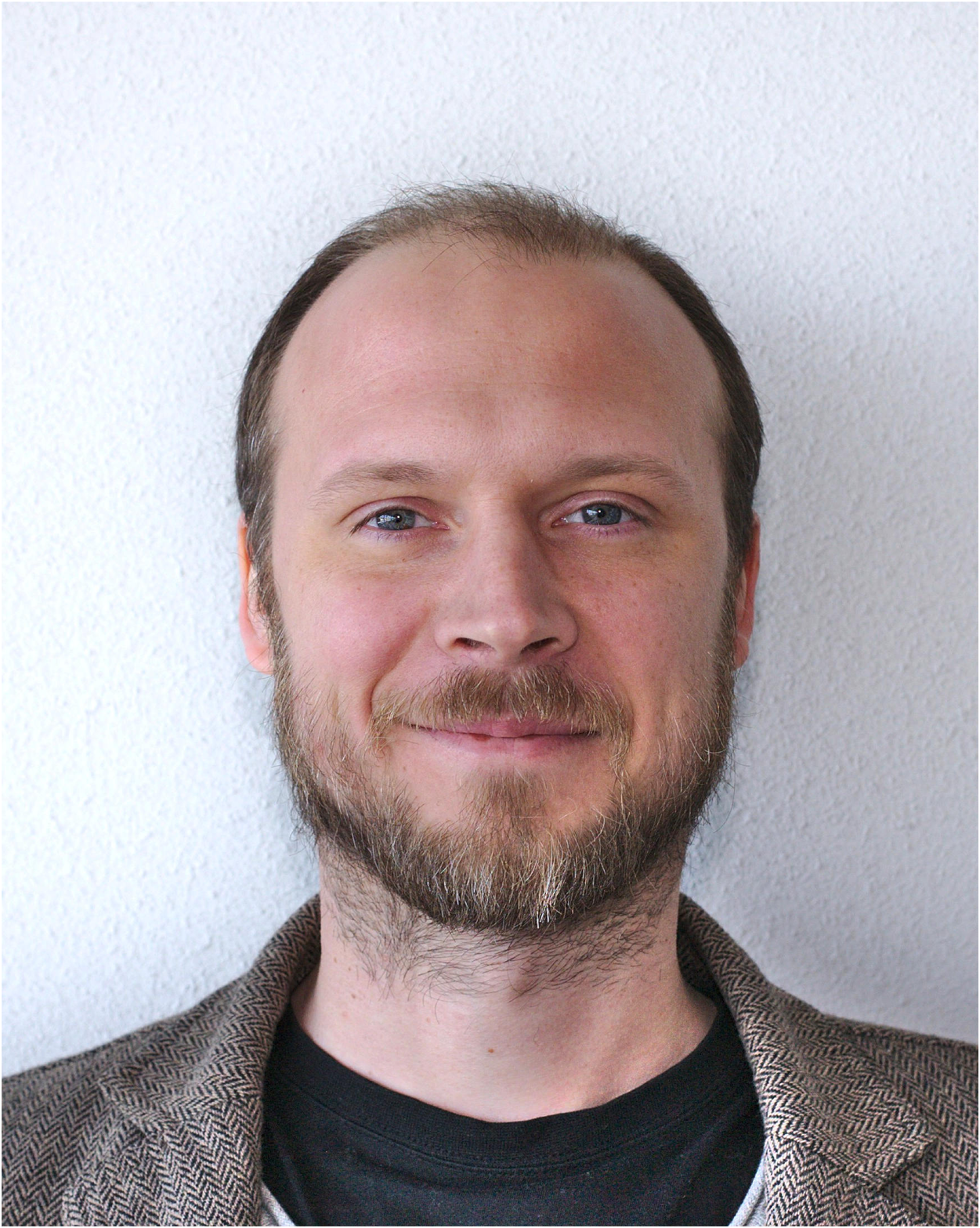}}]
{Alexandre Wagemakers} received the Telecommunication Engineering degree in 2003 from the Polytechnic University in Madrid, Spain; and the Ph.D. degree in Applied Physics in 2008 from the University Rey Juan Carlos, Madrid, Spain, where he is working as an Assistant Professor in the Department of Physics. His interests are currently the theory and applications of Nonlinear Dynamics.
\end{IEEEbiography}

%
%
%




\end{document}